\documentclass[12pt]{article}
\usepackage{amsmath, amsthm, amssymb}

\title{\bf Charge Oscillations in Superconducting Nanodevices Coupled to External Environments}

\author{Robert Alicki$^{a}$, Fabio Benatti$^{b,c}$, 
Roberto Floreanini$^{c}$\\
\small ${}^a$Institute of Theoretical Physics and Astrophyiscs,
University of Gdansk,\\
\small Wita Stwosza 57, 80-952 Gdansk, Poland\\
\small ${}^b$Dipartimento di Fisica Teorica, Universit\`a di Trieste,
Strada Costiera 11, 34014 Trieste, Italy\\
\small ${}^c$Istituto Nazionale di Fisica Nucleare, Sezione di Trieste,
34100 Trieste, Italy}

\date{\null}

\begin{document}

\maketitle

\begin{abstract}
\noindent
Charge oscillations in certain nanodevices, more specifically
the so-called Superconducting Cooper Pair Boxes (SCB), 
are usually interpreted as an effect of macroscopic quantum coherence;
an alternative explanation is however possible in terms of
the Gross-Pitaewski equation for the classical order parameter. 
These two explanations are based on
different quantum states assigned to the SCB, occupation number states 
in the first case, coherent-like states in the second one.
We show that, when the SCB is weakly coupled to an external source of
noise and dissipation, occupation number states are much more
unstable than coherent ones.
\end{abstract}

{\bf 1.} Among the various experimental implementations of qubits, the ones
based on superconducting devices have recently attracted much
interest. In particular, in a series of measures \cite{Naka1,Naka2} it
has been shown that the so-called Superconducting Cooper Pair Boxes (SCB) 
[see section 2 below for a detailed definition] allow 
external control of charge oscillations.
Because of the large number of Cooper pairs involved, these oscillations
have been interpreted as a manifestation of macroscopic quantum 
coherence as theoretically explained by the so-called 
\textit{quantum phase model} \cite{Makhlin,Wendin} which essentially
describes a quantized non-linear oscillator.

A different explanation of the experimental results
is however possible in terms of the solution of a classical 
Gross-Pitaewski equation \cite{A}. In this approach, the charge
oscillations naturally emerge from the time-behaviour of the SCB order 
parameter on the basis of a mean-field treatment.
While the quantum phase model describes the system in terms of
occupation number states,
the ones appearing in the mean-field formulation are close to coherent
states.

Though the predicted frequencies of oscillations in the two models are 
different, the actual accuracy of the experimental data does not allow
to discriminate between them.
Nevertheless, as shown in the following, the two models behave
quite differently in presence of noise induced by a weakly coupled 
external environment.
In such a case, the SCB need to be treated as open quantum 
systems \cite{K,AL,BF,BP}; as a consequence, their dynamics is no
longer unitary, but it is described by a quantum dynamical semigroup,
that takes into account effects of dissipation and decoherence induced by the 
environment.

In this respect, a physically important issue is whether the system is
stable against the presence of noise. It occurs that the two models
both predict relaxation, but with quite different decay properties.
Taking into account the large number of Cooper pairs involved in the
actual experimental implementations of SCB, the decay times of 
occupation number states turn out to be so large that 
charge oscillations would hardly be visible if based on the quantum
phase model; on the other hand, they would be much less affected
by the presence of noise when described in terms of coherent-like
states.
 
This result suggests a possible experimental setting able to
distinguish between the two models of the SCB charge oscillations.
By experimentally inducing a stochastic perturbation of the Josephson 
tunneling, the charge oscillations would be immediately suppressed if
due to macroscopic quantum coherence, while practically unaffected 
otherwise.

\vskip .5cm

{\bf 2.} A SCB consists of a small superconductive 
island, labeled by ``$1$'' in the following, connected
through a Josephson junction to a much larger superconductive island, 
labeled by ``$2$''. 
The islands are inserted in a circuit with a control gate
voltage $V_g$ coupled to them via  a gate capacitor $C_g$: 
in this way one can control the charge on the SCB; further, the total 
number, $N$, of Cooper pairs on the two islands is fixed.

A standard, widely used treatment to model the behaviour of the SCB 
is based on a non-linear oscillator classical Hamiltonian 
\begin{equation}
\label{ham1}
H=4E_C(n-n_g)^2-E_J\cos\theta\ ,
\end{equation}
where $E_C$ is the charging energy of the SCB, $E_J$ the Josephson coupling
energy, $n$ the number of Cooper pairs in island $1$ in excess with
respect to a reference state (with a macroscopic average number of 
Cooper pairs $\overline{n}_1$), $\theta$ is the phase of the
superconducting order parameter of the SCB, while $n_g=C_gV_g/(2e)$ 
accounts for the charging effect due to the control gate 
voltage \cite{Makhlin,Wendin}.

As mentioned in the Introduction, the above Hamiltonian is then
quantized by considering the variable $\theta$ and $n$ as conjugate
phase and number operators satisfying canonical commutation relations.
In the occupation number representation, 
the quantized Hamiltonian takes the form
\begin{equation}
\label{ham2}
\hat{H}=4E_C\sum_{n=0}^N(n-n_g)^2\vert n\rangle\langle
n\vert -\,E_J\,\sum_{n=0}^N\Bigl(\vert n\rangle\langle n+1\vert\,+\,\vert
n+1\rangle\langle n\vert\Bigr)\ .
\end{equation}

In the so called charge qubit regime, the potential barrier between
the two islands is small so that $E_C\gg E_J$. 
By adjusting the control gate voltage so that $n_g=1/2$, one enters a 
particular situation where 
only the states $\vert0\rangle$ and $\vert1\rangle$ play a role and 
are strongly coupled by the Josephson junction.
In such a case, neglecting constant terms, the Hamiltonian (\ref{ham2}) 
can be approximated by a two-level Hamiltonian,
\begin{equation}
\label{ham3}
\hat{H}_{\rm qubit}=-\frac{4E_C(1-2n_g)}{2}\sigma_z-\frac{E_J}{2}\sigma_x\ ,
\end{equation}
with $\sigma_{x,z}$ Pauli matrices, effectively modeling a qubit.
As a consequence, given an initial superposition $|\varphi\rangle$ 
of the eigenstates $\vert 0\rangle$, $\vert 1\rangle$, 
of (\ref{ham3}), the probability of being in the fundamental state at
time $t$, 
\begin{equation}
\label{prob-osc}
p(t):=\left|\langle0\vert{\rm e}^{-it\hat{H}_{\rm qubit}}
\vert\varphi\rangle\right|^2\ ,
\end{equation} 
shows coherent oscillations with frequency $E_J$.
These oscillations have been experimentally
detected \cite{Naka1,Naka2} opening the
possibility to perform qubit operations by controlling the applied gate
voltage and by inserting external magnetic fields.

Observe that an equivalent, more transparent interpretation of this
behaviour of a SCB can be derived within a purely quantum framework
by starting
from a microscopic description in terms of bosonic
annihilation and creation operators
$a_1$, $a_1^\dagger$ and $a_2$, $a^\dagger_2$ of Cooper
pairs in island $1$ and $2$, with fixed
total number $N=a_1^\dagger a_1+a^\dagger_2 a_2$.
Indeed, in a suitable regime, the SCB dynamics can be effectively modelled by the 
Bose-Hubbard Hamiltonian \cite{Cirac,Lewe}
\begin{equation}
\label{ham5}
H_{\rm 2-mode}=E(a_1^\dagger a_1)^2 + U_1\,a_1^\dagger a_1\,+\,U_2\,
a_2^\dagger a_2 - \,K(a_1 a^\dagger_2+a_1^\dagger a_2)\ ,
\end{equation}
where the quadratic term $E(a_1^\dagger a_1)^2$ accounts for
Coulomb repulsion in the small island (the one in the much larger
island $2$ can be neglected);
further, $U_ia^\dagger_i a_i$, $i=1,2$, are potential contributions, while
the last one is a tunneling term.

By substituting the operators $a_i$ with complex numbers
$\sqrt{n_i}{\rm e}^{i\theta_i}$, after some rearrangements and
discarding constant terms, one gets the following classical Hamiltonian
\begin{equation}
\label{ham4}
H=E\Bigl(n_1-\frac{U_2-U_1}{2E}\Bigr)^2-2\,K\,n_1(N-n_1)\cos\theta\ ,
\end{equation}
where $\theta:=\theta_1-\theta_2$.
Let $n_1=n+\overline{n}_1$, with $n$ the number of excess Cooper pairs
relative to the macroscopic average occupation number $\overline{n}_1$ 
in island $1$. 
In SCB experimental applications, $n$ is either $0$ or $1$ and
therefore $n\ll\overline{n}_1$, the average occupation numebr 
$\overline{n}_1$ being typically of
the order of $10^8$.  
One can then approximate $K\, n_1(N-n_1)$ by a constant term 
$K\,\overline{n}_1(N-\overline{n}_1)$ and 
the Hamiltonian (\ref{ham1}) is thus recovered by setting 
$E=4E_C$, $n_g:=(U_2-U_1)/2E-\overline{n}_1$ and 
$E_J=K\,\overline{n}_1(N-\overline{n}_1)$.
The standard qubit interpretation sketched before naturally follows,
whereby the $\sigma_z$ eigenstates $|0\rangle$, $|1\rangle$ really represent
states with a macroscopic number of Cooper pairs. They are indeed orthogonal Fock states,
that differ by one Cooper pair in island ``1''.

\vskip .5cm

{\bf 3.} A different description of the system is possible by treating
the Bose-Hubbard Hamiltonian in a mean-field approach; this is
justified by the large number of Cooper pairs involved.

A generic single Cooper pair state is of the form
\begin{equation}
\label{single-part}
|\psi\rangle=(\psi_1\,a^\dagger_1+\psi_2\,a^\dagger_2)\vert
{\rm vac}\rangle\ ,
\end{equation} 
where $\vert {\rm vac}\rangle$ denotes the vacuum state. 
It is the coherent
superposition of a Cooper pair sitting on island
$1$ and a Cooper pair sitting on island $2$, with amplitudes 
$\psi_i\in\mathbb{C}$ such that $|\psi_1|^2+|\psi_2|^2=1$.
The condensed state of the SCB
is thus appropriately described by the product state
\begin{eqnarray}
\label{MF2a}
|\Psi\rangle_N&:=&
\underbrace{|\psi\rangle\otimes\cdots|\psi\rangle}_{N\,times}
=\frac{1}{\sqrt{N!}}(a^\dagger(\psi))^N|{\rm vac}\rangle\\
\label{MF2b}
&=&\sum_{k=0}^N\sqrt{N\choose
k}\psi_1^k\psi_2^{N-k}|1_{(k)}2_{(N-k)}\rangle\ ,
\end{eqnarray}
where
\begin{equation}
\label{MF2aa}
|1_{(k)}2_{(N-k)}\rangle:=
\frac{(a^\dagger_1)^k(a^\dagger_2)^{N-k}}{\sqrt{k!(N-k)!}}
\vert {\rm vac}\rangle\ ,
\end{equation}
is the occupation number state containing $k$ Cooper pairs in island
$1$ and $N-k$ in island $2$.
More specifically, $\vert\Psi\rangle_N$ embodies the fact that on
islands $1$ and $2$ there is a condensate consisting of a macroscopic 
number $N\gg\overline{n}_1$ of pairs in a same state 
$\vert\psi\rangle$ as in (\ref{single-part}).
Its structure is quite different from the number states 
that are used in the qubit approach: as already remarked at the end
of the previous Section, these correspond
to the occupation number states as in (\ref{MF2aa}) which
contain roughly $\overline{n}_1$ pairs on island
$1$ and $N-\overline{n}_1$ on island $2$.
On the other hand, as shown in (\ref{MF2b}), 
$\vert\Psi\rangle_N$ is a coherent superposition of many occupation
number states on the two islands.

One can easily show that the following relations hold
\begin{eqnarray}
\label{MF5a}
&&
a_1\vert\Psi\rangle_N=\sqrt{N}\psi_1\vert\Psi\rangle_{N-1}\ ,\\
&&
a_2\vert\Psi\rangle_N=\sqrt{N}\psi_2\vert\Psi\rangle_{N-1}\ ;\\
\label{MF5b}
&&
n_1:={}_N\langle\Psi|a^\dagger_1a_1|\Psi\rangle_N=N|\psi_1|^2\ ,\\
&&
n_2:=N-n_1={}_N\langle\Psi|a^\dagger_2a_2|\Psi\rangle_N=N|\psi_2|^2\ .
\end{eqnarray}
The dynamics of $\vert\Psi\rangle_N$ follows the standard
Schr\"odinger equation governed by the Hamiltonian (\ref{ham5})
\begin{equation}
\label{Schr-eq}
i\frac{{\rm d}}{{\rm d}t}\vert\Psi_t\rangle_N=H_{\rm 2-mode}
\vert\Psi_t\rangle_N\ .
\end{equation}
Using (\ref{MF5a}) and (\ref{MF5b}) one finds for the two sides of
this equation
\begin{equation}
\label{MF6}
i\frac{\rm d}{{\rm d}t}\vert\Psi_t\rangle_N=i\sqrt{N}\Bigl(
\dot{\psi}_1(t)\,a^\dagger_1\vert\Psi_t\rangle_{N-1}\,+\,
\dot{\psi}_2(t)\,a^\dagger_2\vert\Psi_t\rangle_{N-1}\ ,
\Bigr)
\end{equation}
while
\begin{eqnarray}
\nonumber
H_{\rm 2-mode}\vert\Psi_t\rangle_N=\!\!\!
&&U_1\sqrt{N}\psi_1(t)\,a^\dagger_1\vert\Psi_t\rangle_{N-1}
+U_2\sqrt{N}\psi_2(t)\,a^\dagger_2\vert\Psi_t\rangle_{N-1}\\
\nonumber
&&-K\sqrt{N}\psi_1(t)\,a^\dagger_2\vert\Psi_t\rangle_{N-1}
-K\sqrt{N}\psi_2(t)\,a^\dagger_1\vert\Psi_t\rangle_{N-1}\\
\label{MF7}
&&+E\sqrt{N}\psi_1(t)\,(a_1^\dagger a_1)a^\dagger_1
\vert{\Psi_t}\rangle_{N-1}\ .
\end{eqnarray}

Substituting  $a^\dagger_1a_1$ in the last line 
with its mean value in (\ref{MF5b}), the
product state $\vert\Psi_t\rangle_N$ is a solution of the evolution equation 
(\ref{Schr-eq})
if the amplitudes $\psi_i(t)$ solve the Gross-Pitaevski equations for
the two component order parameter $(\psi_1,\psi_2)$
\begin{eqnarray}
\label{MF8a}
i\dot{\psi}_1&=&E|\psi_1|^2\psi_1+U_1\psi_1-K\psi_2\\
\label{MF8b}
i\dot{\psi}_2&=&U_2\psi_2-K\psi_1\ .
\end{eqnarray}

By setting $\psi_{1,2}:=\sqrt{n_{1,2}}\, {\rm e}^{i\theta_{1,2}}$ and using the
same definitions as after (\ref{ham4}), in particular
$n_1:=n+\overline{n}_1$ and
$\theta:=\theta_2-\theta_1$, these two equations reduce to 
\begin{equation}
\label{OP3}
\dot{n}=-E_J\sin\theta\,\ ,\quad
\dot{\theta}=\frac{E}{2}\,n
\end{equation}
whence, for small $\theta$, the excess charge number $n$ oscillates
with frequency $\sqrt{(E_JE)/2}$.

As a consequence, the mean-field approach also provides an explanation
for the charge oscillations experimentally observed in SCB, though possibly 
with a different frequency with respect to the one predicted by the
quantum phase model.
However, the experimental setups used so far are unable to measure
frequency oscillations with sufficient accuracy and therefore to
discriminate between the two explanations.

\vskip .5cm

{\bf 4.} From the previous sections, it follows that there are two possible
approaches to describe charge oscillations in SCB, the quantum phase
model which is based on a purely quantum description and the
mean-field one which is semiclassical in nature.
They are both physically consistent with the experimental data; 
although they differ in the prediction of the oscillation
frequencies, this is hardly relevant from the actual experimental
viewpoint.
Instead, we shall now show that
the two approaches give different relaxation patterns 
when the SCB is weakly coupled to an environment which acts as a 
source of noise and
dissipation; this result may indeed have experimental relevance.

We describe the weak coupling of the SCB to an external environment
by means of the total Hamiltonian
\begin{eqnarray}
\label{SCL1}
H_{T}&=& H_E+E(a_1^\dagger a_1)^2\,+\,U_1\,a_1^\dagger a_1\,+\,U_2\,
a_2^\dagger a_2\\
\nonumber
&-&\,K(a_1 a^\dagger_2+a_1^\dagger a_2)
+\lambda\Bigl(
a_1a_2^\dagger \otimes B\,+\,a^\dagger_1a_2 \otimes B^\dagger\Bigr)\ ,
\end{eqnarray}
where $H_E$ is the Hamiltonian of the environment, 
$\lambda\ll1$ is a small coupling
constant and $B$ is a suitable enviroment operator.

We shall assume the environment to be in an equilibrium state and the
corresponding two-point functions to be of
white noise type, $\langle B^\dagger(t)B\rangle_E\simeq\delta(t)$.
Among others, two possible environments exhibit such
behaviour in their correlation functions: a heat bath in which the SCB
is immersed at a temperature very large with respect to the SCB energy 
scales or an external classical stochastic field. 
In particular, the heat bath could consist of non-condensed
electrons in the SCB, whereas the stochastic perturbation can be
practically implemented thorugh a random variation of the Josephson 
tunneling term and therefore be externally controlled.

The delta-like environment correlation functions provide the setting
for the so called \textit{singular coupling limit} \cite{Gorini} 
that leads to a master equation for SCB density
matrices $\rho$ of Kossakowski-Lindblad form:
\begin{eqnarray}
\label{SCL2}
\partial_t\rho&=&-i[H_{\rm 2-mode}+H^{(2)}\,,\,\rho]\,+\,\mathbb{D}[\rho]\\
\nonumber
\mathbb{D}[\rho]&=&
\gamma\Bigl(b\rho b^\dagger-\frac{1}{2}\{b^\dagger
b\,,\,\rho\}\Bigr)+
\delta\Bigl(b^\dagger\rho b-\frac{1}{2}\{bb^\dagger\,,\,\rho\}\Bigr)\\
\label{SCL3}
&& +\beta\Bigl(b\rho b-\frac{1}{2}\{b^2\,,\,\rho\}\Bigr)+
\beta^*\Bigl(b^\dagger\rho
b^\dagger-\frac{1}{2}\{(b^\dagger)^2\,,\,\rho\}\Bigr)\ ,
\end{eqnarray}
where $b:=a_1a_2^\dagger$ and
\begin{eqnarray}
\label{SCL5}
&&
\gamma:=\lambda^2\int_{-\infty}^{+\infty}{\rm d}s
\langle B^\dagger(s)B\rangle_E\ ,\\
&&
\delta:=\lambda^2\int_{-\infty}^{+\infty}{\rm d}s
\langle B(s)B^\dagger\rangle_E\ ,\\
&&
\beta:=\lambda^2\int_{-\infty}^{+\infty}{\rm d}s\langle B(s)B\rangle_E\ . 
\end{eqnarray}
The extra Hamiltonian term $H^{(2)}$ is a bath induced correction and
the coefficients $\beta\in\mathbb{C}$, $\gamma\geq 0$ and
$\delta\geq0$ satisfy the condition 
$\gamma\delta\geq|\beta|^2$ which ensures the complete positivity 
\cite{K,AL,BF} of
the one-parameter quantum dynamical semigroup generated by
(\ref{SCL2}).
It incorporates the dissipative effects on the dynamics
of the SCB due to the presence of an external environment.

It is enough to limit our considerations to the study of
the stability properties of any initial
SCB pure state $\vert \phi\rangle\langle \phi\vert$. 
We shall thus focus on the decay constant at time $t=0$, to which only
the term $\mathbb{D}[\rho]$ in (\ref{SCL2}) contributes,
\begin{eqnarray}
\nonumber
&&
\Gamma_{\phi}:=
-\langle \phi|\mathbb{D}[|\phi\rangle\langle\phi|]|\phi\rangle=
\gamma\Bigl(|\langle \phi|b|\phi\rangle|^2-\langle \phi|b^\dagger 
b|\phi\rangle\Bigr)
+\delta\Bigl(|\langle \phi|b|\phi\rangle|^2-\langle 
\phi|bb^\dagger|\phi\rangle\Bigr)\\
\label{SCL6}
&&\hskip 7cm +2\Re\Bigl(\beta\Bigl((\langle \phi|b|\phi\rangle)^2
-\langle \phi|b^2|\phi\rangle\Bigr)\Bigr)\ ,
\end{eqnarray}
where $\Re$ stands for real part; $\Gamma_\phi$ measures the velocity 
with which any state 
$\vert\phi\rangle\langle\phi\vert$ initially departs from itself. 

We shall show that the decay constant $\Gamma_\phi$ strongly depends on
whether the initial state of the SCB is an occupation number (Fock)
or a coherent-like state. We stress again that the first choice underlies
the widely accepted qubit interpretation of the charge oscillations, while the other
is at the root of the mean-field approach.

\medskip

\noindent
$\bullet$ {\bf Quantum phase model}

As already mentioned at the end of Section 2, the states that play a role in the qubit 
explanation of SCB oscillations are occupation number states. 
Thus, we shall consier as initial state $|\phi\rangle$ the 
eigenstate $|0\rangle$ of $\sigma_z$, which is of the form
$\vert \phi\rangle=\vert1_{(n_1)}2_{(N-n_1)}\rangle$, with \hbox{$n_1\simeq \bar n_1\simeq 10^8$}.
By observing that
\begin{eqnarray}
\label{DC1a}
&&\hskip -.7cm
b\vert 1_{(n_1)}2_{(N-n_1)}\rangle
=a_1a_2^\dagger\vert 1_{(n_1)}2_{(N-n_1)}\rangle
=\sqrt{n_1(N-n_1+1)}
\vert1_{(n_1-1)}2_{(N-n_1+1)}\rangle\\
\label{DC1b}
&&\hskip -.7cm
b^\dagger\vert 1_{(n_1)}2_{(N-n_1)}\rangle
=a_1^\dagger a_2\vert 1_{(n_1)}2_{(N-n_1)}\rangle
=\sqrt{(n_1+1)(N-n_1)}\vert1_{(n_1+1)}2_{(N-n_1-1)}\rangle\\
\label{DC1c}
&&\hskip -.7cm
b^\dagger b\vert 1_{(n_1)}2_{(N-n_1)}\rangle
=n_1(N-n_1+1)
\vert 1_{(n_1)}2_{(N-n_1)}\rangle\\
\label{DC11a}
&&\hskip -.7cm
bb^\dagger\vert 1_{(n_1)}2_{(N-n_1)}\rangle
=(n_1+1)(N-n_1)
\vert 1_{(n_1)}2_{(N-n_1)}\rangle\ \
\end{eqnarray}
the decay constant is computed to be
\begin{equation}
\label{cavsco2}
\Gamma_{\rm qubit}=\gamma n_1(N-n_1+1)
+\delta(n_1+1)(N-n_1)\ .
\end{equation}
\medskip

\noindent
$\bullet$ {\bf Mean-field model} 

The states involved in this approach are instead of the form
$\vert\Psi\rangle_N$ given in (\ref{MF2a}).
Using relations (\ref{DC1a})--(\ref{DC11a}), one can show that 
\begin{eqnarray}
\label{DC2a}
&&
_N\langle\Psi\vert b\vert\Psi\rangle_N
=\sqrt{n_1(N-n_1)}{\rm e}^{i\theta}\\
&&
\label{DC2b}
_N\langle\Psi\vert b^\dagger b\vert\Psi\rangle_N
=n_1\Bigl(N-n_1+\frac{n_1}{N}\Bigr)\\
&&
\label{DC2c}
_N\langle\Psi\vert bb^\dagger \vert\Psi\rangle_N
=(N-n_1)\Bigl(1+n_1-\frac{n_1}{N}\Bigr)\ ,
\end{eqnarray}
so that $\vert\Psi\rangle_N$ indeed behaves like a coherent state for the
operator $b$, with labeling 
parameter $\sqrt{n_1(N-n_1)}\,\exp(i\theta)$.

With the help of the previous relations, the decay constant in the
mean-field model explicitly reads
\begin{equation}
\label{cavsco3}
\Gamma_{{\rm mean-field}}
=\frac{n_1^2}{N}\gamma+\delta(N-n_1)(1-\frac{n_1}{N})
-2\Re\Bigl(\beta{\rm e}^{2i\theta}\Bigr)\frac{n_1(N-n_1)}{N}\ .
\end{equation}

In typical experimental conditions, the average number $n_1$ of Cooper pairs
in island $1$ is essentially the macroscopic occupation number 
$\overline{n}_1\simeq 10^8$, while the total number $N$ of Cooper pairs in both
islands is such that $\overline{n}_1\ll N$. 
Taking this into account and comparing the two decay constants 
in (\ref{cavsco2}) and (\ref{cavsco3}), one finds
\begin{equation}
\label{final} 
\frac{\Gamma_{\rm qubit}}{\Gamma_{\rm mean-field}}\simeq \overline{n}_1\ .
\end{equation}
As a consequence, because of the large value of $\overline{n}_1$, 
the two decay constants turn out to be very different in magnitude. 

In presence of an external source of
noise and dissipation, occupation number states in the quantum phase
model seem to be much more unstable than the semiclassical, coherent
like states in the mean-field approach.
This gives the possibility of distinguishing the two descriptions.
What looks experimentally implementable is a setup in which an
externally controlled stochastic white noise is injected into the
SCB device, for instance by suitably modifying the Josephson
junction characteristic parameters and thus the potential barrier.
By switching on such a stochastic perturbation, charge oscillations  
should be suppressed if due to macroscopic quantum coherence as
described by the quantum phase model; instead, if they survive, that
would be an indication of a semiclassical origin as modeled by the
mean field approach.


\begin{thebibliography}{99}

\bibitem{Naka1}
Y. Nakamura, Yu. Pashin and J.S. Tsai, Nature \textbf{398} (1999) 786

\bibitem{Naka2}
Y. Nakamura, Yu.A. Pashkin and J.S. Tsai, Phys. Rev. Lett. \textbf{87}
(2001) 246601

\bibitem{Makhlin}
Y. Makhlin, G. Sch\"on, A. Schnirman, Rev. Mod. Phys. \textbf{73} (2001) 357

\bibitem{Wendin}
G. Wendin, V.S. Schumeiko, Superconducting Quantum Circuits, Qubits and Computing,
cond-mat/0508729

\bibitem{A}
R. Alicki, Quantumness of Josephson Junctions Reexamined, quant-ph/0610008

\bibitem{K}
V. Gorini, A. Frigerio, M. Verri, A. Kossakowski and E.G.C. Sudarshan,
Rep. Math. Phys. {\bf 13} (1978) 149

\bibitem{AL}
R. Alicki and K. Lendi, {\it Quantum Dynamical Semigroups and 
Applications}, Lect. Notes Phys. {\bf 286}, (Springer-Verlag, Berlin, 1987)

\bibitem{BF}
F. Benatti, R. Floreanini, Int. J. Mod. Phys. B \textbf{19}
(2005) 3063

\bibitem{BP}
H.-P. Breuer and F. Petruccione, {\it The Theory of Open
Quantum Systems} (Oxford University Press, Oxford, 2002)

\bibitem{Cirac}
D. Jaksch, C. Binder, J.I. Cirac, C.W. Gardiner and P. Zoller,
Phys. Rev. Lett. \textbf{81} (1998) 3108

\bibitem{Lewe}
M. Lewenstein, A. Sanpera, V. Ahufinger, B. Damski, A. Sen De nd U. Sen,
Adv. in Phys. \textbf{56}  (2007) 243

\bibitem{Gorini}
V. Gorini, A. Kossakowski, J. Math. Phys. \textbf{17} (1976) 1298
\end{thebibliography}
\end{document}